\documentstyle[11pt,aaspp4,flushrt]{article}
\def\Rin{R_{\rm in}}
\def\Rms{R_{\rm ms}}
\def\Rout{R_{\rm out}}

\def\Rg{R_{\rm g}}
\def\Ka{{\rm K}{\alpha}}

\def\ginga{{\it Ginga}\ }

\def\dof{\,{\rm dof}}
\def\Firr{F_{\rm irr}}

\def\LX{L_{\rm X}}
\def\rtr{r_{\rm tr}}
\def\mdcrit{{\dot m}_{\rm crit}}
\righthead{Accretion in Nova Muscae 1991}
\lefthead{\.{Z}ycki, Done, \&  Smith}

\begin{document}

\title{Evolution of the accretion flow in Nova Muscae 1991}

\author{Piotr T. \.{Z}ycki and Chris Done}
\affil{Department of Physics, University of Durham, South Road,
Durham DH1 3LE, England; piotr.zycki@durham.ac.uk, chris.done@durham.ac.uk}
 
\and
 
\author{David A. Smith}
\affil{Department of Physics and Astronomy, University of Leicester,
       University Road, Leicester LE1 7RH, England; das@star.le.ac.uk}
 
\begin{abstract}

We identify the X--ray reflected component in the \ginga spectra of 
Nova Muscae 1991, 
a Black Hole transient system used as the prototype for the recent
model of \markcite{es97}Esin, McClintock \& Narayan (1997) 
based on advection dominated  disk solutions. We see that the reflected 
spectrum is generally  significantly relativistically  smeared,
and use this, together with the amplitude of reflection, to track the innermost
extent of the accretion disk. The optically thick disk switches from
being highly ionized to nearly neutral during the transition from high to low
state, and the inner radius of the disk moves outwards during the low state
decline. 

Qualitatively, this overall trend is compatible with Esin et al.'s model,
but quantitatively, the retreat of the inner disk during the high to low
state transition is much slower than predicted. The hard (low state)
spectra are not produced solely by an optically thin accretion flow: optically
thick material within $\sim 20-100\, \Rg$ is generally present. 

\end{abstract}

 
\section{Introduction}

Black hole binary systems give one of the most direct ways in which to study 
the
physics of accretion disks. There is no surface boundary layer or strong 
central
magnetic field to disrupt the flow, and the orbital parameters are often well
studied so that the inclination, mass and distance of the system are tightly
constrained. Additionally, many of these systems (the Soft X--ray Transients,
hereafter SXT) show dramatic outbursts where the luminosity rises rapidly 
from
a very faint quiescent state to one which is close to the Eddington limit, 
and then
declines again over a period of months, giving a clear sequence of spectra as a
function of mass accretion rate. The usual pattern for such objects is for the
outburst to be dominated by a soft component of temperature $\sim 1$ keV, with
or without a (rather steep) power law tail, while the later stages of the
decline show much harder power law spectra, extending out to 100--200 keV
(see e.g. \markcite{ts96}Tanaka \& Shibazaki 1996). 

The ``standard'' accretion disk model developed by \markcite{ss73}Shakura \& 
Sunyaev (1973;
hereafter SS) derives the accretion flow structure in the limit when the
gravitational energy released is radiated locally in a geometrically thin disk.
Such models give temperatures of order 1 keV for SXT in outburst, but are
unable to explain the presence of a (hard or soft) power law tail to high
energies. Either there are parts of the disk flow in a different configuration
to that of SS, or there is some non--disk structure such as a corona powered by
magnetic reconnection (e.g.\ \markcite{hmg94}Haardt, Maraschi \& Ghisellini 
1994).

Recently, another solution of the accretion flow was postulated to
explain the hard X--ray data. Below a
critical accretion rate, $\mdcrit$, a stable, hot, optically thin,
geometrically thick solution can be found if radial energy transport 
(advection)
is included (see e.g.\ the review by \markcite{n97}Narayan 1997 and the 
discussion by \markcite{ch95}Chen et al.\ 1995 of
how all the accretion flow solutions fit together). In the
absence of an optically thick disk
the hot electrons can only cool via cyclotron/synchrotron
emission (on an internally generated magnetic field), bremsstrahlung, or
Compton scattering of the resultant spectra of these two processes. 
The lack of strong flux of soft photons in these models means that the X--ray 
spectra are typically hard. Such flows were proposed to explain the hard and 
very faint X--ray spectra seen from black holes  candidates (BHC) in
quiescence (\markcite{nmcy}Narayan, McClintock \& Yi 1996). These models 
were then extended by \markcite{es97}Esin, McClintock \& Narayan (1997; 
hereafter EMN) to cover the whole range of
luminosity seen in SXT. EMN assume that accretion takes place via an SS disk
for the incoming material beyond some (large) truncation radius of $\rtr\sim
2\times 10^4 \Rg$ (where $\Rg=GM/c^2$), but that interior to this the flow 
{\it always}\/ jumps to the advective solution if the latter exists.  
During the outburst  the mass accretion rate goes above that at which an 
advective flow can be  sustained,
so they assume that the disk adopts the SS solution, forming a cool flow down 
to $\Rms\equiv 6\, \Rg$ and hence producing soft spectra. During the 
decline the mass accretion rate falls below $\dot{m}_{\rm crit}$, producing 
an intermediate stage as the SS
disk retreats. The transition radius between the advective and SS flow
progressively increases, and the low state proper is defined to be where the SS
disk is again restricted to beyond $2\times 10^4\, \Rg$, although the spectrum
 is
indistinguishable from this for transition radii $\ge 200\, \Rg$.  This change
from an SS to advective disk solution takes place over a very small range in
total luminosity, and is postulated to explain the soft--hard transition seen 
in
the SXT.  EMN tested their model on \ginga data of Nova Muscae 1991, and
found that it could indeed reproduce the overall spectral evolution of the
source.

This model predicts large changes in the accretion geometry as a function of
mass accretion rate (see Figure 1 of EMN) which can be observationally tested
by spectral studies of X--ray reprocessing. X--rays interact with optically 
thick
material to produce a reflected continuum due to Compton scattering and the 
iron $\Ka$ line due to fluorescence/recombination.  Their normalizations 
determine the solid angle of the reprocessor as seen from the X--ray 
source while the shape  of
the reflected continuum below 10 keV, the line energy and its equivalent width 
are functions of ionization state of the reprocessor. If the reflecting
material is indeed a disk extending close to the black hole then
the reprocessed spectrum is strongly smeared by the combination of high orbital
velocities and strong gravity (\markcite{fa89}Fabian et al.\ 1989; 
\markcite{la91}Laor 1991; \markcite{ro96}Ross, Fabian \&  Brandt 1996). 
In the EMN model it is clear that 
as the source changes from high to low state then the
amount of reflection should decrease, together with the amount of relativistic
smearing, as the optically thick reflecting disk moves outwards.

In this Letter we present results of spectral analysis of \ginga data of soft
X--ray transient GS~1124-68 (Nova Muscae 1991) with the emphasis on detailed
modeling of the effects of X--ray reprocessing. Our results confirm the
prediction of EMN that the covering fraction of the reprocessing matter
decreases as the source luminosity declines. However, the normalization of the
reflected component and the amount of relativistic smearing are both
significantly larger than the model predictions, indicating that the SS disk
retreats outwards more slowly than predicted during the intermediate and low
states.  The hard (low state) spectra seen in the persistent BHC and SXT during
decline are incompatible with the EMN scenario of an optically thin 
accretion flow extending out to $\sim 10^4\Rg$, and instead require that 
there is a transition from an optically
thin to optically thick structure at $\sim 20-100\, \Rg$.  

\section{Data}

Observations of Nova Muscae 1991 and various aspects of data modeling were
described in detail by \markcite{eb94}Ebisawa et al.\ (1994).  
We re--extracted the data from the \ginga archive in Leicester and subtracted
the background using a modified version (\.{Z}ycki, Done \& Smith 1997) of 
standard procedures (\markcite{ha89}Hayashida et al.\ 1989). For our analysis 
we have selected six data sets: January 11th, May 18th, June 13th, June 21st,
July 23rd, and September 3rd.  These are the only datasets in which the hard 
tail
is sufficiently developed to search for reflection, and where
the instrument is accurately pointed (within $0.3^\circ-0.4^\circ$) at 
the source. 

\section{Model for the hard X--ray emission}

\label{sec:model}
Our basic model of the X--ray spectrum is a sum of a primary power law and
the reprocessed component which is a sum of the Compton--reflected
continuum and the iron $\Ka$ line near $6.5\,$keV. The reflected continuum
is computed using the angle--dependent Green functions of
\markcite{mz95}Magdziarz \& Zdziarski (1995), with the possibility of
partial ionization (parameterised by ionization parameter, $\xi\equiv\LX/n
r^2$) as in \markcite{do92}Done et al.\ (1992), all as implemented in the
``pexriv'' model in {\sc XSPEC} version 9.01 (\markcite{ka96}Arnaud 1996). 
This is a very basic ionization code, balancing the photo--ionization from
the power law irradiation with radiative recombination at a given fixed
temperature, which we set to $10^6$ K. The self--consistent iron $\Ka$
line for this ionization state, computed using the Monte Carlo code
described in \markcite{zc94}\.{Z}ycki \& Czerny (1994), is then added to
the continuum. 

The relativistic effects expected from an accretion disk around a
(non--rotating) black hole are simulated by convolving this total (line
plus continuum) spectrum with the {\sc XSPEC} ``diskline'' model. This is
parameterised by the inner and outer radius of the disk, $\Rin$ and
$\Rout$, its inclination and the irradiation emissivity, $\Firr(r)$. The
latter three are fixed at $10^4\,\Rg$, $60^\circ$ (\markcite{or96}Orosz et
al.\ 1996) and $\propto r^{-3}$, respectively, while we fit for $\Rin$,
the ionization state $\xi$, and the amplitude of the reprocessed component
expressed as a solid angle subtended by the reprocessor from the X--ray
source, $\Omega$, normalized to $2\pi$, i.e.\ $f\equiv\Omega/2\pi$ (see
also \markcite{zds97}\.{Z}ycki et al.\ 1997). 

Our code also allows the ionization to vary as a function of radius,
$\xi(r)\propto r^{\beta}$, by splitting the disk into radial rings of
constant ionization state. The relative covering fraction of each ring is
determined by the emissivity and the relativistic corrections are given by
the extent of each ring. 

\section{Model fitting}

During its decline phase Nova Muscae 1991 went through a sequence of
distinct ``states'' (Ebisawa et al.\ 1994; EMN).
During and soon after the outburst the source was in
the Very High State (VHS; January 11th data), with both a strong soft 
component
and a power law tail. Two and a half months later that changed into High State
(HS), where the soft component dominates, and further one  and a half  months
later, in mid-May, the Intermediate State (IS) began in which the soft 
component
dramatically decreases while the power law tail hardens and increases in
intensity. This is then followed two months later by the Low State (LS; 
July 23rd and September 3rd data sets), where the hard power law dominates
 the luminosity 
output. An important point for the January (VHS) and May (HS/IS) data is the 
description
of the soft thermal component since it dominates the spectrum below $\sim
4\, $keV. We model this by optically thick comptonization of soft seed photons
(\markcite{ti94}Titarchuk 1994; as implemented in the ``comptt'' model 
in {\sc XSPEC}), 
since simple blackbody and disk blackbody models (with or without relativistic 
corrections) cannot give an adequate description of the observed spectral 
shape  (see also \markcite{g97b}Gierli\'{n}ski et al.\ 1997b). 
The results for the June and July data
are not sensitive to a choice of the soft component so here we use the
multi-temperature disk blackbody model (\markcite{mi84}Mitsuda et al.\ 1984).

First, we have checked that the spectral features seen in the data at 
5--8 keV, which we will attribute to X--ray reprocessing, cannot be accounted
for by the effects of Comptonization, even if a broad distribution of soft seed
photons is assumed. To this end we have modelled the May 18th data using
``comptt'' for the soft component and we convolved it with the relativistic
Green functions of \markcite{ti94}Titarchuk (1994) to obtain the model
for the hard component. Such a two component model fails to account for 
the data
($\chi^2=656/25\dof$) and the residuals clearly indicate the presence of
the reprocessed component. The situation is not improved if the soft component
is modeled using the Green functions of \markcite{st80}Sunyaev \& Titarchuk 
(1980) convolved with a blackbody.

\begin{figure*}
 \plotone[50 440 600 700]{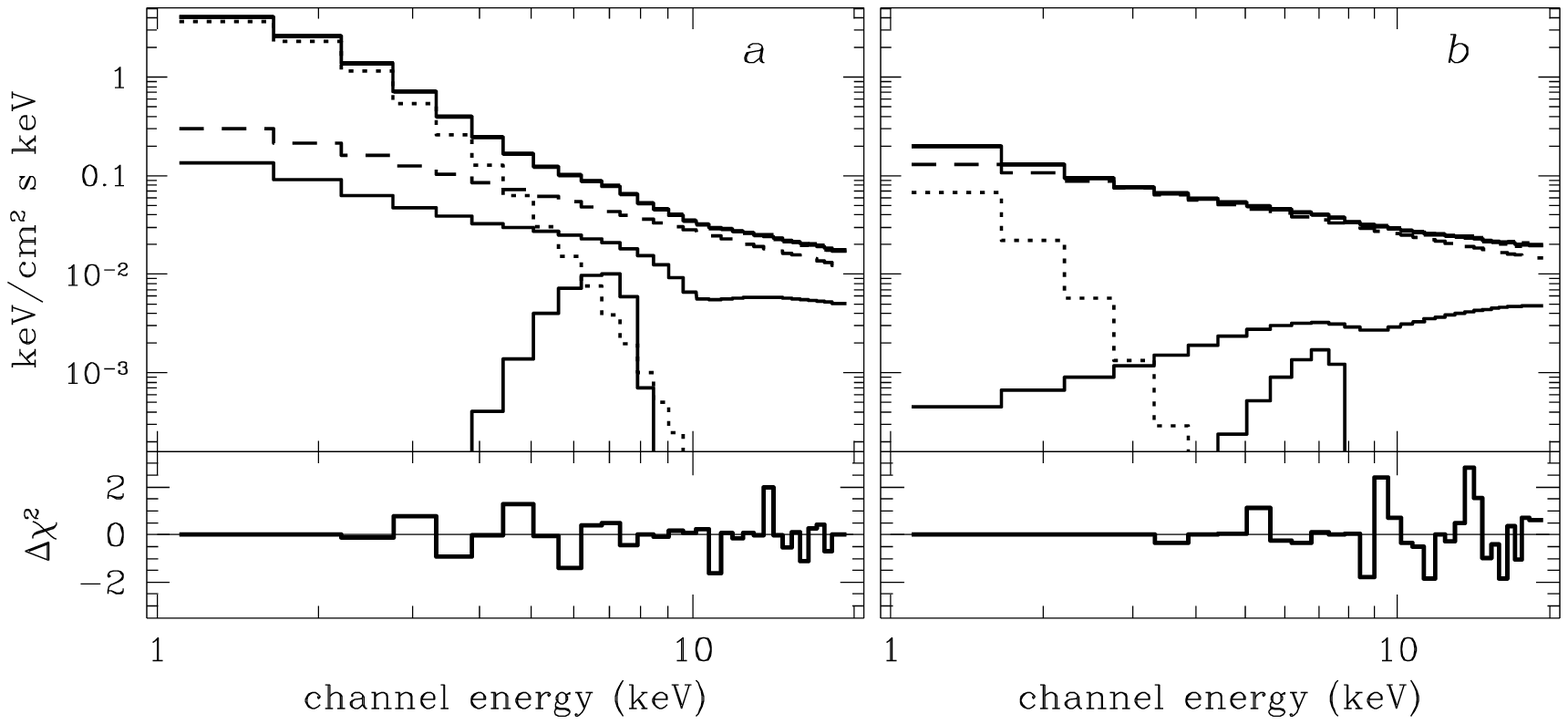}
\figcaption{Best fit spectra for May 18th data ({\it a}\/) and June 13th data 
({\it b}\/) and the fit residuals.
The upper solid histograms are the total spectra, the dotted histograms
show the soft thermal components and the dashed histograms show the primary
power law spectrum
The shape of the reprocessed component (lower solid histograms for the 
reflected continuum and the Fe line), determined mainly by ionization
parameter $\xi$, changed dramatically but its relative amplitude, $f$, 
and amount of relativistic smearing
changed only slightly (see Table~\ref{tab:inter}) indicating only moderate
change of the inner disk radius. \label{fig:spectra}}
 \end{figure*}

We first model the reprocessed component assuming its uniform ionization
($\xi(r)={\rm const}$; Table~\ref{tab:inter}). 
Reflection is significantly detected in all but
the September 3rd dataset, where the signal to noise is insufficient to obtain
interesting constraints. Even in the VHS, where the soft component is very 
strong, reflection is significantly detected and its amount is quite well
determined. It is strongly ionized and relativistically smeared
(see \markcite{g97b}Gierli\'{n}ski et al.\ 1997b for a similar result 
for the soft  state of Cyg X--1). 
At the beginning of the IS (May 18th data set) the reflector is still
highly ionized, but as the importance of the soft component declines, the
ionization state also drops (to $\xi \sim 10$ where iron is predominantly 
ionized  only to Fe
VII--IX), and the power law hardens by $\Delta\Gamma\sim 0.4$ without
a significant change in its total flux. Relativistic smearing of the reflection
spectrum is significant in all the datasets except for July 23rd and 
September 3rd. The best--fit spectra for the May 18th and June 13th data sets 
are shown in Figure~\ref{fig:spectra}.

Figure~\ref{fig:cont} shows the
derived confidence contours for $\Rin$ and $f$ during the IS and LS.  
Clearly the data
require that the solid angle decreases as the source declines, and while
there is no statistical {\it requirement\/} that the relativistic smearing 
decreases, the data are consistent with a monotonic increase in $\Rin$. 

To see how these results depend on any ionization structure, we
repeated the fits for $\xi(r)\propto r^{-1.5}$ and $r^{-4.5}$, as expected 
from a
gas and radiation pressure dominated SS disk, respectively (with the assumed
$\Firr\propto r^{-3}$). Only the contours
for the May 18th dataset change, since this is the only one shown which has
strong ionization. We include these results in Figure~\ref{fig:cont} 
by the dashed and dotted
lines for the gas and radiation pressure dominated cases, respectively.  Both
allow much larger reflected fractions, since there is now a contribution from
very highly ionized material at the inner disk edge which gives negligible
spectral features. While this material is also strongly affected by the
relativistic smearing, the lack of features means that there are no
observational signatures until further out in the disk where the ionization
drops, so allowing smaller values of $\Rin$.

We have also checked that a different model for the soft component
in the May data does not change the results concering the reprocessed
component. Using the solution
of \markcite{st80}Sunyaev \& Titarchuk (1980) convolved with a blackbody
spectrum to model the soft component we re-derived  the confidence contours
in the $f$ vs.\ $\Rin$ plane. The best fit and minimum values of $f$ and
$\Rin$ agree to within 10\% with previous results (solid curve in 
Fig.~\ref{fig:cont}) while the maximum values are larger by 20\% when
the Sunyaev \& Titarchuk solution is used.

\section{Discussion}

We can use the results of our spectral fitting to derive the geometry, and then
compare this with that proposed by EMN in their scenario based on advective 
disk.  Our
derived limits on the inner radius of the accretion disk from the amount of
relativistic smearing are dependent on the illumination emissivity,
for which we assumed $\Firr\propto r^{-3}$. This should be a good 
approximation  for a central quasi--spherical source even if
there is some overlap between the cold disk and hot source 
(\markcite{pkr97}Poutanen, Krolik \& Ryde 1997). Thus it seems likely that 
the radii derived in the previous section are reasonable unless
the disk flares substantially, and/or there are significant non--local effects
in the energy generation and/or relativistic corrections are important. 

In the VHS, HS and beginning of the IS, where the spectrum contains a strong 
soft component, the EMN model predicts that the disk extends down to the last 
stable orbit at $6 \Rg$. From the fits to the May (HS/IS) data we see that 
indeed  the reflector subtends a large  solid angle with respect to 
the hard X--ray  source and  the spectrum is strongly relativistically 
smeared, although the data prefer an inner radius of $\sim 20\, \Rg$, and are
formally inconsistent with $6\,\Rg$. Relativistic corrections to the
illumination emissivity will be important at these small radii, although the
results of \markcite{rb97}Reynolds \& Begelman (1997) suggest that this 
should lead to a
steeper dependence, rather than the flatter one needed for the data to be
consistent with $6\, \Rg$. A more likely possibility to retrieve a disk down to
the last stable orbit is that the strong ionization of the reflector seen in
this state is not adequately modeled (the ionization balance derived from 
\markcite{do92}Done et al.\
1992 is rather approximate), and that this has distorted the fit. This strong
ionization of the upper layers of the disk probably results from the intense
gravitational energy release within the SS flow, and it may lead to
Comptonization of the emerging soft
photons, producing the observed spectral shape of the thermal emission
(\markcite{rf93}Ross \& Fabian 1993; \markcite{g97b}Gierli\'{n}ski et al.\ 
1997b).

\begin{figure}
 \plotone{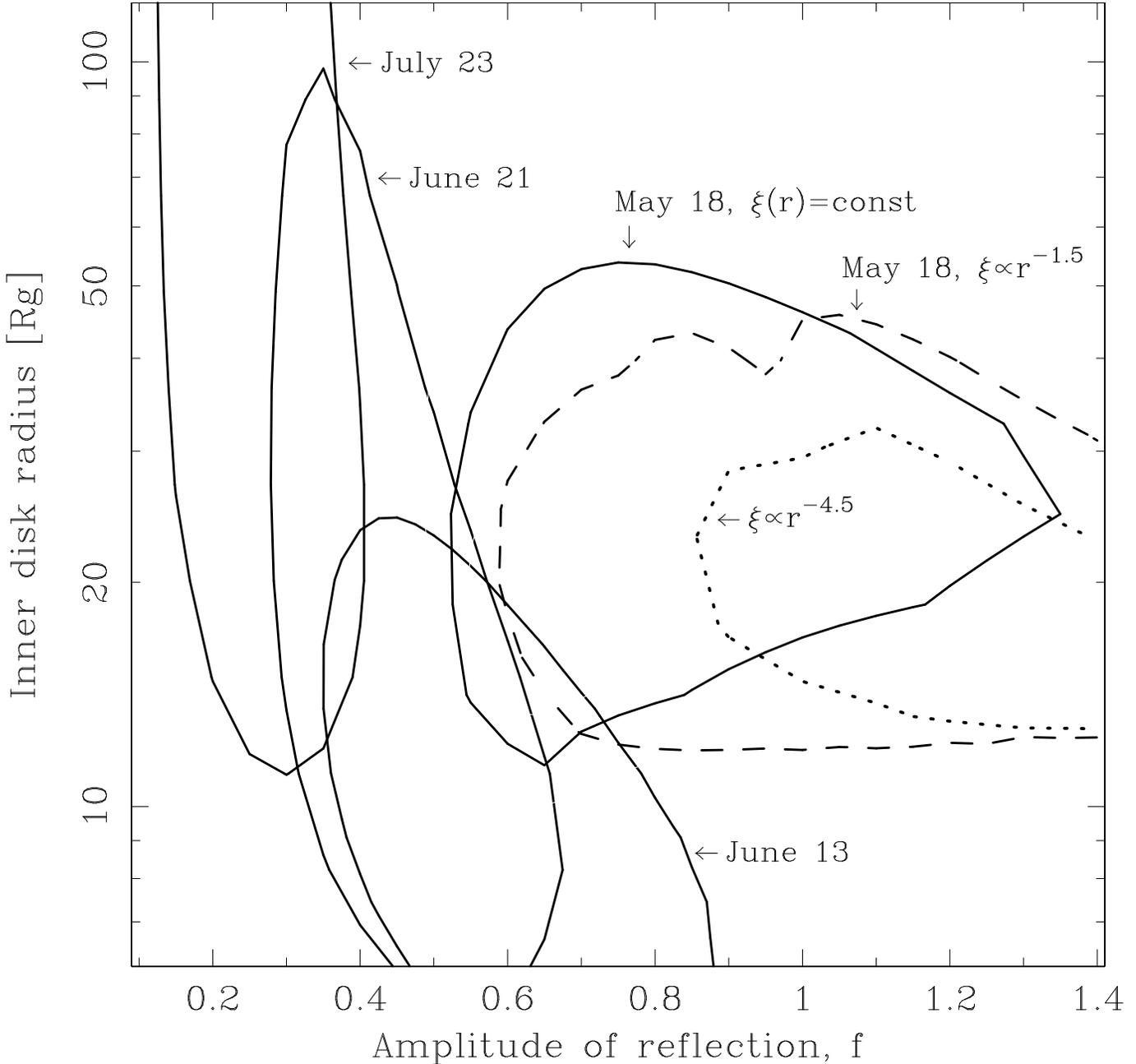}
 \figcaption{Contours of $\Delta\chi^2=4.61$ 
(90\%  confidence level for two interesting parameters)
as a function of the amplitude of the
reprocessed component, $f$, and the inner radius of the disk, $\Rin$,
as derived from spectral modelling of the data.
Solid contours are for constant ionization parameter models
while the dashed and dotted contours for the May 18th data are for 
models with $\xi(r)\propto r^{-1.5}$ and 
$\xi(r)\propto r^{-4.5}$, respectively (see \S~\ref{sec:model}).
The results are consistent with a monotonic decrease of $f$ and increase
of $\Rin$ indicating the retreat of the inner disk as the decline phase
was progressing. \label{fig:cont} }
\end{figure}

However, the next stage of evolution in the IS does not seem to be properly
described by the EMN model. Since the hard X--ray flux was roughly constant
between May and June the observed dramatic decrease of the ionization, from
$\xi\sim 10^4$ to $\sim 10$, must be related to the disappearance of the soft
component. Consequently, the power law becomes harder, since the availability 
of seed photons for Compton cooling is dramatically reduced.  However, the 
dramatic
spectral change was {\it not\/} accompanied by any significant change of 
$\Rin$, although the best fit value of $f$ decreases by $\sim 0.2 - 0.3$.
This could be produced by a small increase in $\Rin$, from $\sim 10\, \Rg$ to
$\sim 20\, \Rg$, depending on the emissivity, which is within the error 
contours
of Figure~\ref{fig:cont}.  It is clearly incompatible with the prediction of
EMN that $\Rin$ should be $\sim 200\,\Rg$ for the June data since this would
produce no significant relativistic smearing of the iron spectral features and
too small amplitude of reflection. 

Similarly, the properties of the reprocessed component in the July data (LS) 
are incompatible with the EMN computations. While the effect of smearing 
constrains $\Rin$
to be $> 20\, \Rg$, the amplitude of reflection, $f\sim 0.25$, requires $\Rin$ 
to be $\sim 30\, \Rg$ for the EMN geometry (obtained using simple Monte
Carlo code for photons propagation without further interactions after their
generation). With $\Rin\approx 10^4\,\Rg$, 
as postulated by EMN, the amplitude $f$ would be negligible.  
The values of $f$ inferred by us are also in agreement with those for
Cyg X-1 and GX 339-4 in low states, from both simple fits to the data and 
detailed  modeling of the continuum emission
(\markcite{ue94}Ueda, Ebisawa \& Done 1994; 
\markcite{eb96}Ebisawa et al.\ 1996;
\markcite{g97a}Gierli\'{n}ski et al.\ 1997a;  
\markcite{do97}Dove et al.\ 1997;  
\markcite{pkr97}Poutanen et al.\ 1997).
The amplitude $f\sim 0.2-0.3$ seems thus to be a typical value for 
BHC in the low state, and shows that concept of accretion occurring in a
optically thin flow from very large radii ($\sim 10^4\Rg$) cannot be sustained.

Thus, the observed transition from high (soft) to low (hard) state does seem to
involve a retreat of the optically thick material, as in the EMN model.
However, the inner disk radius is much smaller than the values postulated by
EMN in their intermediate and low states.  Since their treatment of the
transition radius marking the edge of the advective flow and the standard SS
disk is assumed rather than calculated then this may not be an insurmountable
problem. The major spectral change in their model occurs as $\Rin$ changes
from $6$ to $\sim 30\Rg$, which corresponds well to the observed constraints 
during the
high/low state transition, suggesting that it is indeed associated with the
innermost regions of the accretion flow becoming optically thin.  However, the
sudden switching of {\it all\/} the accretion flow into an optically thin state
over a very small range in $\dot{m}$ is a major facet of their model, and is 
clearly inconsistent with the observed behavior of BHC in general and Nova
Muscae in particular.

\acknowledgements
 
This research
made use of data obtained from the Leicester Database and Archive
Service at the Department of Physics and Astronomy, Leicester University.
C.D.\ acknowledges support from a PPARC Advanced Fellowship.

%\clearpage

\clearpage

\clearpage

\clearpage

\begin{deluxetable}{lccccc}
 \footnotesize
 \tablewidth{0pt}
 \tablecaption{Results of model fitting, $\xi(r)={\rm const}$ 
 \label{tab:inter}}
 \tablehead{
 \colhead{ data }  & \colhead{$\Gamma$\tablenotemark{a}} & 
                     \colhead{$f$\tablenotemark{b}} &
                     \colhead{$\xi$\tablenotemark{c} }  & 
                     \colhead{$\Rin$\tablenotemark{d}\ ($\Rg$)} & 
\colhead{$\chi^2/\dof$ (spect.\ bins)}
}
\startdata
Jan 11, VHS   & $ 2.03\pm 0.16$           &    $0.30^{+0.13}_{-0.04} $  & 
         $(1.0^{+1.0}_{-0.7})\times 10^4 $ & 
         $13^{+5}_{-3}$        & 26.1/31 (40) \nl
May 18, HS/IS & $ 2.29^{+0.05}_{-0.03} $  &    $ 0.64^{+0.40}_{-0.10} $  & 
         $ (3.5^{+9.5}_{-3.0})\times 10^4$ & $18^{+22}_{-6} $ & 
                                                          $13.5/22$ (31) \nl
June 13, IS & $1.91\pm 0.03$  &  $0.57_{-0.17}^{+0.23}$  &
        $10^{+22}_{-8} $  &  $10^{+9}_{-4}$ &      $20.9/24$ (31) \nl
June 21, IS & $1.83\pm 0.03$ &  $0.46\pm 0.12$  &  
        $ 4^{+12}_{-3}$ & $13^{+25}_{-6}$    &       $11.9/24$ (31) \nl
July 23, LS &  $1.72\pm 0.02 $  &  $0.24_{-0.08}^{+0.11} $  &
        $17^{+40}_{-16} $  &  $50^{+\infty}_{-35}$   & $15.3/24$ (31) \nl
Sept 3, LS  & $1.95\pm 0.11 $   &  $0^{+1.1}$ &
     0(f)  & -- & 26.9/26 (31) \nl
\enddata
\tablenotetext{a}{Photon spectral index}
\tablenotetext{b}{Covering fraction of the reprocessor}
\tablenotetext{c}{Ionization parameter of the reprocessor}
\tablenotetext{d}{Inner radius od the accretion disk}

\end{deluxetable}

\end{document}